\documentstyle[aps]{revtex}
\begin{document}

   \title{Magnetic Radius of the Deuteron}
   \vskip 0.2 in
 \author{Andrei Afanasev\thanks{On leave from Kharkov Institute of Physics and
Technology, Kharkov 310108, Ukraine}\\ 
{\normalsize\it North Carolina Central University, Durham, NC 27707, USA}\\ 
{\normalsize\it and}\\
{\normalsize\it Jefferson Lab, Newport News, VA 23606, USA}\\ 
\bigskip
 V.D. Afanasev and S.V. Trubnikov\\
 {\normalsize\it Kharkov State University, Kharkov 310077, Ukraine}} 
\maketitle

\begin{abstract}
The root-mean square radius of the deuteron magnetic moment 
distribution, $ r_{Md}$,  is calculated for several realistic models of
the $NN$--interaction. For the Paris potential the result is
$r_{Md} = 2.312 \pm 0.010 \;$ fm. The dependence of $r_{Md}\;$ on the
choice of $NN$ model, relativistic effects and meson exchange currents
is investigated. The experimental value of $r_{Md}$ is also considered. 
The necessity of new precise measurements of the deuteron magnetic 
form factor  at low values of $Q^2$ is stressed.
  \end{abstract}

\begin{sloppypar}

The root-mean-square radius (RMSR) $r_{Md}\;$ of the magnetic moment 
spatial distribution in the deuteron is defined in the usual way as: 
 \begin{equation}
r_{Md}\equiv <r_{Md} ^2>^{1/2} =\biggl[- \frac{6G'_{Md}(Q^2=0)}
{G_{Md}(Q^2=0)} \biggr]^{1/2} =\biggl[-\frac{3G'_{Md}(0)}{\mu _d}
\biggr]^{1/2}\;,
\label{RMSR}
\end{equation}
where  $G_{Md}(Q^2)\;$ is the deuteron magnetic form factor (DMFF),
$Q^2 $ is the modulus of the four-momentum transfer squared, and $ \mu _d $
is the deuteron magnetic moment. The radius $ r_{Md} $ is an independent
static property of the deuteron, which is not directly connected to the
deuteron charge  radius $r_{Cd} $. Here we see a clear difference
between deuteron and nucleon structure. Indeed,
if we assume that the scaling law for the nucleon FF's
 \begin{equation}
G_{Ep}(Q^2)=\frac{G_{Mp}(Q^2)}{ \mu _p}= \frac{G_{Mn}(Q^2)}{\mu _n}
\label{scale}
\end{equation}
is valid for very low values of $ Q^2 $ (and we have serious experimental
and, especially, theoretical reasons to think so), then the immediate result
from eq.~(\ref{scale}) for the three nucleonic radii is
 \begin{equation}
  <r^2_{Ep}> = <r^2_{Mp}>= <r^2_{Mn}> \:,
\label{Mradii}
\end{equation}
i.e. the charge and magnetic radii of the proton coincide. In the 
deuteron case, the theoretical foundations for a scaling law between the charge
DFF and magnetic one are absent, so the theoretical values of $ r_{Cd} $
and $ r_{Md} $ must be considered independent. 

Let us first discuss the experimental status of $ r_{Md}$. The results of
measurements of $ G_{Md}(Q^2) $ for low $ Q^2 \alt 1 $ fm$^{-2} $
are contained in refs.\cite{Gold} - \cite{Jones}. These experimental values
were approximated by a polynomial of  degree $ n $:
 \begin{equation}
 G_{Md}= 2\mu _d \biggl[1-\frac{1}{6}<r^2_{Md}>\cdot Q^2 +\sum_{p=2}^n
a_p Q^{2p}\biggr] \:.
\label{GMd}
\end{equation}
The optimal order of the polynomial in eq.~(\ref{GMd})
appears to be $ n=2$. In this way we obtained an experimental value 
 \begin{equation}
r_{Md}= 1.90 \pm 0.14 \;\;\mbox{ fm}.
\label{expRMd}
\end{equation}
Note that from experiments on elastic electron--deuteron $ (ed) $ 
scattering, the deuteron charge radius $ r_{Cd} $ is determined 
much better than $ r_{Md} $ in eq.~(\ref{expRMd}). The best 
analysis \cite{Sick}  of experimental data leads to the following 
result: 
 \begin{equation}
 r_{Cd}=  2.128 \pm 0.011\;\; \mbox{fm}.
\label{expRCd}
\end{equation}
One should also consider the value based on atomic isotope shift 
measurements of H\"ansch et al. \cite{Pach} which give a value 
$ r_{Cd}=  2.136 \pm 0.005 $ \mbox{fm} \cite{FMS}.
Comparing eqs.~(\ref{expRMd}) and (\ref{expRCd}), we see 
that the two radii are approximately equal:  $ r_{Md} 
\approx r_{Cd}\:,$ but the precision of the determination of $ 
r_{Md} $ is low as a consequence of the low precision of the available 
experimental data. 

Now let us turn to the theoretical calculation of $ r_{Md} $. The standard
expression in the literature 
\footnote{We have several concerns about the correct formula for $ 
G_{Md}(Q^2) $ in NRIA and plan to discuss it separately. The possible 
modification of eq.~(\ref{DMFF}) would not influence any 
calculations in the static limit $ Q^2 \rightarrow 0$.} 
for the DMFF in the non-relativistic impulse approximation (NRIA) is
 \begin{equation}
G_{Md}(Q^2)=\frac{M_d}{M}\biggl[2G^S_{MN}(Q^2)\cdot C_S(Q^2) +
G^S_{EN}(Q^2)\cdot C_L(Q^2) \biggl]\;,
\label{DMFF}
\end{equation}
where
 \begin{eqnarray}
C_S(Q^2)&=& \int_0^{\infty} {[ u^2(r) -\frac{1}{2}w^2(r)]
j_0(\frac{1}{2}Qr) + \frac{1}{\sqrt{2}} w(r)[u(r) +
\frac{1}{\sqrt{2}} w(r)]j_2(\frac{1}{2}Qr)}dr\;,\nonumber \\
C_L(Q^2)&=& \frac{3}{2}\int_0^{\infty} w^2[j_0(\frac{1}{2}Qr) +
j_2(\frac{1}{2}Qr) ]dr\;. \nonumber
\end{eqnarray}
In eq.~(\ref{DMFF}), $ u(r), w(r) $ are the deuteron radial  S--, D-- 
state wave-functions; $ j_{0,2} $ are the spherical Bessel functions; 
$ G^S_{.\;N}\equiv \frac{1}{2}(G_{.\;p} + G_{.\;n}) $ are the 
isoscalar nucleon FFs; $ M_d\:,M $ are the  deuteron and nucleon 
masses, respectively. {}From eqs.~(\ref{RMSR}),~(\ref{DMFF}) we have 
 \begin{eqnarray}
<r^2_{Md}>&=&\frac{1}{\mu _d}\cdot\Biggl\{ \bigl[<r^2_{Mp}> +
<r^2_{Mn}>\bigr] (1-\frac{3}{2}p_d)+  \nonumber\\
&&+(\mu _n + \mu _p)\cdot\int^\infty _0 \bigl( \frac{1}{4}\cdot u^2 -
\frac{1}{10\sqrt 2}u\cdot w - \frac{7}{40}w^2 \bigr)r^2\cdot dr +\nonumber\\
&& +\frac{3}{4}p_d\cdot\bigl[ <r^2_{Ep}> + <r^2_{En}>\bigr] +
\frac{9}{80}\int_0^\infty w^2\cdot r^2 dr \Biggr\}\:,
\label{RMSR2}
\end{eqnarray}
where, as usual,  $p_d = \int_0^\infty w^2(r)dr$.

The results of calculations of $r_{Md} $, following eq.~(\ref{RMSR2}),
are listed in the third column of the Table~\ref{Tab} for several
realistic deuteron wave functions (the NN potentials are identified
in the first column). Due to lack of experimental information about the
magnetic radii of the proton and, especially, the neutron, 
eq.~(\ref{scale}) was used. For charge radii of the proton and neutron the
inputs are the well-known experimental values $ <r^2_{Ep}>^{1/2}=0.862\pm
0.012\; \mbox{fm \cite{Rp} and }<r^2_{En}> =-0.1194\pm 0.0018\; \mbox{fm}^2$.
More recent measurements \cite{Harvey} give a consensus value of  
$<r^2_{En}> =-0.1140\pm 0.0026\; \mbox{fm}^2$.

Before discussing these results, we should estimate the
contributions to $r_{Md} $ from relativistic effects (RE) and meson
exchange currents (MEC). It's reasonable to think that for $Q^2 \rightarrow
0 $ the functions which describe the contributions  of RE and MEC are
small and smooth enough, so that their first derivatives will be negligible.

Indeed, for RE, in the framework of the
formalism~\cite{Gour} the formula for the DMFF appears to be 
\footnote{As a results of our own calculations, it seems that the 
relativistic formula for $G_{Md} $ in ref.~\cite{Gour} has several 
inaccuracies. Here we have corrected it.} 
 \begin{equation}
 G_{Md}= \frac{M_d}{M}\biggl[
\frac{2G^S_{MN}}{\sqrt{1+\tau }}\cdot C_s+ \bigl(G^S_{EN} +
\frac{\tau}{1+\sqrt{1+\tau }} G^S_{MN}\bigr)\cdot \frac{C}{1+\tau }\biggr]
\;, \label{DMFF_corr} 
\end{equation} 
where 
 \begin{eqnarray*} C&=&C_L+
C_1,\\ C_1&=&\frac{9}{2}\int_0^\infty dr\cdot w^2(r)\cdot \frac{1}{Qr}
\int_0^{Qr/2}j_2(\alpha )d\alpha \:,\\
\tau &=&Q^2/4M^2
\end{eqnarray*}
and the other notations were introduced in eq.~(\ref{DMFF}). The RE in
eq.~(\ref{DMFF_corr}) are either of Darwin--Foldy (DF) nature (the factor
$1/\sqrt{1+\tau }$), or due to nucleon motion (NM) in the deuteron.
{}From eq.~(\ref{DMFF_corr}) we obtain
 \begin{equation}
<r^2_{Md}>=<r^2_{Md}>_{NRIA} + <\Delta r^2_{Md}>_{RE}\:,
\label{RMDRE}
\end{equation}
 \begin{eqnarray*}
<\Delta r^2_{Md}>_{RE}&=&<\Delta r^2_{Md}>_{DF}+<\Delta r^2_{Md}>_{NM}=\\
&=&\frac{3}{4M^2}-\frac{9}{16M^2}\cdot p_d (\frac{\mu _n+\mu _p+2}{\mu_d})\:,
\end{eqnarray*}
where the first term in eq.~(\ref{RMDRE}) is given by eq.~(\ref{RMSR2}). 
The contribution of RE to the value of $r_{Md} $ is 
shown in Table~\ref{Tab} in the fourth column. 

As far as the MEC are concerned, it's well-known that the general 
formalism describing these effects is rather complicated. For an 
estimate we use the simple parametrization of the MEC contribution to 
$G_{Md}$ for low $Q^2$, which was given in~\cite{Lomon}: 
 \begin{equation}
\bigl( \Delta G_{Md} \bigr)_{MEC} =\beta _1 e^{-\alpha _1Q^2}\:,
\label{MEC}
\end{equation}
so the correction to $r_{Md} $ is
\[\bigl( \Delta r^2_{Md} \bigr)_{MEC} = 3\frac{\alpha _1\beta _1}{\mu
_d}\:.\]
For the Reid soft core potential (RSC), $\beta _1= 0.0288 \;\;\mbox{and}
\;\; \alpha _1= 0.16 \;\mbox{fm}^2 $ and the result is also given in the
Table ~\ref{Tab}. No doubt, we could introduce and analyse more
refined versions of RE and MEC contributions, but 
as they are small it may not be necessary. 

The main result is evident: the agreement between the calculated
theoretical and experimental values of $r_{Md} $ is poor. 

So let us discuss in more detail what we have learned. 
Accepting that the theoretical expressions for $r_{Md} $ are 
reliable, the main contributions to $ r_{Md} $ emerge from the first 
and second terms in eq.~(\ref{RMSR2}).  The contributions of the 
last two terms and the other degrees of freedom (additional to NRIA) are 
comparable and small. So the theoretical value of $r_{Md} $ depends 
mainly on the D--state probability in the deuteron, and on the 
magnetic radii of the neutron and proton. In particular, a small 
deviation from the scaling law (eqs.~(\ref{scale}), (\ref{Mradii})) 
may produce a variation in $r_{Md}$, which is comparable to that 
following from a variation of $p_d$. In this sense it will be 
very interesting to have results from a direct measurement of the 
neutron magnetic  radius $ r^2_{Mn} $ in experiments on 
neutron--electron scattering (as was done for the determination of $ 
<r^2_{En}> $ in experiments on thermal neutron scattering from atomic 
electrons). Once the experimental values of $r_{Mp}\; \mbox{and}\; 
r_{Mn} $ are tied down, the theoretical value of $r_{Md} $ will 
depend mainly on the value of $p_d$. 

Now we make some comments about the determination of experimental 
value of $ r_{Md} $. The present status of low--$Q^2 $ experiments on 
elastic ed--scattering at large angles $ \Theta _e \sim 180^\circ $ is 
such that we cannot extract the value of $ r_{Md} $ from experimental 
data with sufficient (at least as compared to eq.~(\ref{expRCd})) 
accuracy. Indeed, for low values of $ Q^2 $  there are only five 
experimental points of $ G_{Md} $ \cite{Gold}--\cite{Jones}. These 
values were obtained using the old generation of electron 
accelerators, and the experimental errors in $ G_{Md}(Q^2) $ are 
large. For comparison it may be noted that for determination of the 
deuteron charge radius many more experimental points were 
used; moreover for low $ Q^2 $ the longitudinal part $ A(Q^2) $ of 
elastic ed--scattering was measured with high accuracy\cite{Platch}. 
So for determination of $ r_{Md} $ new precise detailed measurements 
of $ G_{Md}(Q^2 \rightarrow 0) $ are very desirable. 

An improved (as compared to eq.~(\ref{expRMd})) experimental value of 
$r_{Md} $ will be useful in the following directions.  Firstly, for 
inclusion of $ r_{Md} $ to the standard set of static deuteron 
properties when investigating a realistic model  of the 
nucleon--nucleon interaction. Secondly, a knowledge of the exact 
value of $ r_{Md} $ is necessary for calculations of the hyperfine 
splitting of the electronic levels in deuterium (corrections  due to 
the finite size spatial distribution of the deuteron magnetic moment: 
see, for example, ref.~\cite{Azam}). The hyperfine splitting is the 
good example of the intersection of high and low energy physics. In 
principle we may invert the problem and try to extract the value of $ 
r_{Md} $  from experimental data on splitting of the atomic S--levels 
in deuterium. Lastly, $ r_{Md} $ will remain the main piece of 
information about the magnetic distribution of the neutron, until 
direct measurements of this quantity in neutron--electron scattering 
experiments are realized. 
\end{sloppypar} 
\acknowledgements
We would like to thank Donald Sprung for useful discussions. 
The work of A.A. was partially supported by the US Department of Energy under
contract DE--AC05--84ER40150.

 \begin {table}
\caption{Magnetic radius of the deuteron $r_{Md}$ (in fm)}
\begin{tabular}{|lccc|}
                 &$p_d   $& nonrelativistic & with relativistic \\
NN potentials    & (\%) &  $ r_{Md} $     &  corrections  \\
                 &        &    (eq. 8)     &  (eq. 10)   \\
\tableline
Bonn \cite{Mach} &  4.25  &2.346 $\pm$ 0.010\tablenote{The errors
in the other listed values of $r_{Md}$, due to errors in
the nucleon radii, are the same.} & 2.353 $\pm$ 0.010 \\
Paris \cite{Lacomb} &  5.77  &2.312              & 2.318        \\
Nijmegen \cite{Nag} &  5.92  &2.314              & 2.321        \\
Reid (RSC) \cite{Reid}\tablenote{For RSC inclusion of the
MEC correction leads to the result $r_{Md} = 2.305 \pm0.010\;$fm.}
                         &  6.46  &2.295  & 2.302             \\
Moscow State     &&&                                          \\
University \cite{Kuku}  &  6.74  & 2.292  & 2.299          \\
                 &&&                                 \\
\tableline
\multicolumn{4}{|c|}{Experimental value is $r_{Md}=1.90\:\pm0.14\;$fm.}
\end{tabular}
\label{Tab}
\end{table}

\end{document}